\documentclass[conference]{IEEEtran}
\IEEEoverridecommandlockouts
\usepackage[noadjust]{cite}
\usepackage{amsmath,amssymb,amsfonts}
\usepackage{algorithmic}
\usepackage{graphicx}
\usepackage{textcomp}
\usepackage{wasysym}
\usepackage{romannum}
\usepackage{booktabs}
\usepackage{xcolor}
\usepackage{url}
\def\BibTeX{{\rm B\kern-.05em{\sc i\kern-.025em b}\kern-.08em
    T\kern-.1667em\lower.7ex\hbox{E}\kern-.125emX}}
\begin{document}

\title{ Two-Stage Distributed Energy Resources Scheduling via Chance-Constrained AC Optimal Power Flow: A Second-Order Cone Programming Approach
{\footnotesize }
}

\author{\IEEEauthorblockN{Mingyue He, \textit{Student Member}, \textit{IEEE}, Zahra Soltani, \textit{Student Member}, \textit{IEEE}, and Mojdeh Khorsand, \textit{Member}, \textit{IEEE}}
\IEEEauthorblockA{School of Electrical, Computer and Energy Engineering \\
Arizona State University, Tempe, AZ, USA\\
Email: mingyue1@asu.edu, zsoltani@asu.edu, mojdeh.khorsand@asu.edu}
}

\maketitle

\begin{abstract}
The penetration of distributed energy resources (DERs) is increasing dramatically. Due to the uncertainty of DERs, the operation of the distribution system is facing higher risks and challenges. To overcome such challenges, a two-stage chance-constrained convex AC optimal power flow (ACOPF) model is proposed in this paper, which can increase the economic efficiency of distribution system operation and manage the intermittency of DERs. In the first stage, a convex second-order cone programming (SOCP)-based ACOPF model is proposed in which the detailed models and limitations of DER, namely, demand response (DR), energy storage units, and rooftop PV systems are modeled to obtain participation ratio of DERs. In the second stage, Monte Carlo simulation is utilized to model the uncertainties of DERs. A probability violation index is introduced to make a trade-off between scheduling more DERs and imposing a higher risk to the distribution system. In this stage, power flow analysis is conducted for each scenario to determine the probability violation index of system. Then, a modified SOCP-based ACOPF is proposed to satisfy the system probability violation criterion. Simulation results illustrate that the proposed two-stage chance-constrained model improves economic efficiency and reliability of real-time operation of the distribution system.
\end{abstract}

\begin{IEEEkeywords}
AC optimal power flow (ACOPF), distributed energy resources, chance-constraint, stochastic optimization, convex relaxation.
\end{IEEEkeywords}

\section*{Nomenclature}
\addcontentsline{toc}{section}{Nomenclature}
\noindent
\textit{Sets}
\noindent
\begin{IEEEdescription}[\IEEEusemathlabelsep\IEEEsetlabelwidth{$P_{n,t}^{pv3},Q_{n}^{pv3}$}]
\item[$\phi_{b}$] Set of bus number.
\item[$\phi_D,\phi_D(n)$] Set of bus number of DR, and set of DR connected to bus $n$.
\item[$\phi_{F},\phi_{F}(n)$] Set of bus number of controllable capacitor device, and set of controllable capacitor devices connected to bus $n$.
\item[$\phi_{G},\phi_{G}(n)$] Set of bus number of substation, and set of substations connected to bus $n$.
\item[$\phi_n(n)$] Set of buses connected to bus $n$.
\item[$\phi_{S},\phi_{S}(n)$] Set of bus number of storage, and set of storage connected to bus $n$.
\item[$\phi_{1},\phi_{1}(n)$] Set of bus number of PV type 1, and set of PV type 1 connected to bus $n$.
\item[$\phi_{2},\phi_{2}(n)$] Set of bus number of PV type 2, and set of PV type 2 connected to bus $n$.
\item[$\phi_{3},\phi_{3}(n)$] Set of bus number of PV type 3, and set of PV type 3 connected to bus $n$.
\item[$L$] Set of line from bus $n$ and to bus $i$.
\end{IEEEdescription}
\noindent
\textit{Parameters}
\begin{IEEEdescription}[\IEEEusemathlabelsep\IEEEsetlabelwidth{$P_{n,t}^{pv3},Q_{n}^{pv3}$}]
\item[$ \rho^{W},\rho^{D},\rho^{R}$] The wholesale, PV unit, and DR electricity price.
\item[$B_{n,i},G_{n,i}$] Susceptance and Conductance of line $(n,i)$.
\item[$C_{n,i}^{l}$] Capacity limit of line $(n,i)$.
\item[$C_{n}^{P1},C_{n}^{S1}$] Active and apparent power capacity limit for PV type 1.
\item[$C_{n}^{P2}$] Active power capacity limit for PV type 2.
\item[$C_{n}^{P3},C_{n}^{S3}$] Active power and apparent power capacity limit for PV type 3.
\item[$C_{n}^{SH},C_{n}^{SL}$] Maximum and minimum state of charge of the battery storage.
\item[$D_{n}^P,D_{n}^Q$] Active and reactive power demand at bus $n$.
\item[$F_{n}^{H}$] Controllable capacitor device limit.
\item[$N_{S}$] The number of scenarios generated by MCS.
\item[$R_{n}^{H},R_{n}^{L}$] Maximum and minimum battery output.
\end{IEEEdescription}
\noindent
\textit{Decision Variables}
\begin{IEEEdescription}[\IEEEusemathlabelsep\IEEEsetlabelwidth{$P_{n,t}^{pv3},Q_{n}^{pv3}$}]
\item[$P_{n}^{g},Q_{n}^{g}$] Active and reactive power from the grid.
\item[$P_{n,i}^{l},Q_{n,i}^{l}$] Active and reactive power flow on line $(n,i)$.
\item[$P_{n}^S,SE_{n}$] Output power, state of charge.
\item[$PSE_{n}$] available of each battery storage.
\item[$P_{n}^{DR},Q_{n}^{DR}$] Active and reactive power from DR.
\item[$P_{n}^{pv1},Q_{n}^{pv1}$] Active and reactive power of each PV type 1.
\item[$P_{n}^{pv2}$] Active power for each PV type 2.
\item[$P_{n}^{pv3},Q_{n}^{pv3}$] Active and reactive power of each PV type 3.
\item[$Q_{n}^{CC}$] Controllable capacitor device reactive power.
\item[$I_{n,i},R_{n,i},U_{n}$] Auxiliary variables.
\end{IEEEdescription}

\section{Introduction}

The distribution system operations progressively rely more on the energy coming from the distributed energy resources (DERs). However, this huge evolution in the distribution grid poses some challenges in the operation paradigms due to the inherent characteristics of DERs, namely uncertainty, and variability. Such uncertainties are imposed due to the human-in-the-loop effects and fluctuations of renewable resources. The inadequate management of uncertainty and variability of scheduled DERs can undermine the reliability and security of the power system and lead to higher operating costs. 

Many published papers have addressed the uncertainty modeling to overcome the aforementioned challenges \cite{Reference5}-\cite{Reference12}. A probabilistic method called chance-constrained is proposed for addressing uncertainty in decision-making problems \cite{Reference5}-\cite{Reference6}. The chance-constrained based optimization models can search the feasible region to yield a solution that can guarantee a determined level of risk. Reference \cite{Reference5} and \cite{Reference6} present a comprehensive review of the three main chance-constraint categories and provide numerical simulations for the chance-constraint based DC optimal power flow (OPF) model. However, DERs are scattered in the distribution grid. DCOPF model is not suitable for the distribution system due to the high R/X ratio and losses. A chance-constrained based convex ACOPF using semidefnite programming (SDP) is proposed in \cite{Reference21}. The authors introduce a penalty term on power losses to obtain the near-global optimal solution of ACOPF. However, the authors only consider the forecasted error and fail to consider the characteristics of different types of DERs in their system. The authors in \cite{Reference13} present a chance-constraint based convex ACOPF optimization model for the transmission system and introduce the reshaped confidence region for the robustness of their solution. The robust optimization-based method is one main category of chance-constraint. A robust optimization-based model with convex second-order cone programming (SOCP)-based ACOPF is proposed in \cite{Reference23}. Multiples robust optimization-based models are proposed to deal with the uncertainty of the DERs \cite{Reference22}-\cite{Reference12}. However, the robust optimization may not be computationally tractable if the uncertainty set is poorly designed.

In this paper, a two-stage chance-constrained SOCP-based ACOPF model is proposed to tackle the uncertainty of DERs in the balanced distribution grid. The proposed chance-constrained based algorithm provides the solution that can guarantee the risk of the system is within the determined value. In the first stage, a SOCP-based ACOPF model is proposed to identify the optimal dispatch of DERs and power purchasement from the wholesale market, and obtain the participation ratio of DERs in the system. A probability violation index is introduced to make a trade-off between scheduling more DERs and imposing a higher risk on the distribution system. In the second stage, uncertainties of DERs are modeled by Monte Carlo simulation (MCS) and a power flow analysis is conducted for each scenario of MCS to determine the probability violation index of the system. Then, a modified SOCP-based ACOPF is proposed to satisfy the system probability violation criterion.
 
The rest of the paper is organized as follows. Section \Romannum{2} introduces proposed SOCP-based ACOPF with DERs. The proposed chance-constrained model is presented in Section \Romannum{3}. The simulation results and conclusions are shown in Section \Romannum{4} and Section \Romannum{5}, respectively.

\section{Proposed SOCP-based ACOPF Model}
The proposed SOCP-based ACOPF model with integration of DERs is presented in this section. The DERs include demand response (DR), energy storage units, and rooftop PV systems. The objective function of the proposed model which is expressed in Eq. (\ref{Obj}) is enhancing the economic efficiency of the distribution system by minimizing the operating cost of purchasing power from the grid, cost of PV units type 1, 2 and 3, which are explained later in this section, and cost of incentivizing customers for participating in DR.
\begin{flalign}
    \label{Obj}&\min \sum_{n \in \phi_{G}}  \rho^{W} P_{n}^{g} + \sum_{n \in \phi_{1}} \rho^{D} P_{n}^{pv1} + \sum_{n \in \phi_{2}} \rho^{D} P_{n}^{pv2}  + && \nonumber \\
    & \qquad \sum_{n \in \phi_{3}} \rho^{D} P_{n}^{pv3} +  \sum_{n \in \phi_D} \rho^{R} P_{n}^{DR} &&
\end{flalign}

In this paper, a SOCP-based ACOPF model with integration of DERs is proposed to convexify the non-convex ACOPF model by introducing three auxiliary variables, i.e., $R_{n,i}$, $I_{n,i}$ and $U_{n}$ in (\ref{I})-(\ref{U}) \cite{Reference1}. 
\begin{flalign}
    \label{I} &R_{n,i} = - V_n V_i \cos\theta_{n,i} && \forall (n,i)\in L \\
    \label{R} &I_{n,i} = - V_n V_i \sin\theta_{n,i} && \forall (n,i)\in L \\
    \label{U} &U_{n} = \frac{V_n}{\sqrt{2}}  && \forall n \in \phi_{b}
\end{flalign}
where $V_n$ and $V_i$ are the voltage magnitude at bus $n$ and $j$. The $\theta_{n,i}$ is the angle difference between bus $n$ and bus $j$. The SOCP- based active and reactive line flow constraints are shown in (\ref{line_P})-(\ref{Line_Cone}). The line flow limit constraint is given by (\ref{Line_limit}).
\begin{flalign}
    \label{line_P}&P_{n,i}^{l} = \sqrt{2} G_{n,i} U_{n} - G_{n,i} R_{n,i} + B_{n,i} I_{n,i} && \forall (n,i)\in L \\
    \label{line_Q}&Q_{n,i}^{l} = \sqrt{2} B_{n,i} U_{n} - B_{n,i} R_{n,i} - G_{n,i} I_{n,i} &&  \forall (n,i)\in L \\
    \label{Line_Cone}&2U_{n} U_{i} \ge R_{n,i}^2 + I_{n,i}^2 && \forall (n,i)\in L\\
    \label{Line_limit}& - C_{n,i}^{l} \le P_{n,i}^{l} \le C_{n,i}^{l} && \forall (n,i)\in L
\end{flalign}

The node active and reactive power balance constraints of the system are expressed in (\ref{nodeBalance_P}) and (\ref{nodeBalance_Q}), respectively.
\begin{flalign}
    \label{nodeBalance_P}&\sum_{n \in \phi_{G}(n)} P_{n}^{g} + \sum_{n \in \phi_D(n)} P_{n}^{DR}+ \sum_{n \in \phi_{1}(n)} P_{n}^{pv1} + \sum_{n \in \phi_{2}(n)} P_{n}^{pv2} \nonumber \\
    &+ \sum_{n \in \phi_{3}(n)} P_{n}^{pv3} + \sum_{n \in \phi_{S}(n)} P_{n}^{S} = D_{n}^P + \sum_{j \in \phi_n(n)} P_{n,i}^{l} \\
    \label{nodeBalance_Q}&\sum_{n \in \phi_{G}(n)} Q_{n}^{g} + \sum_{n \in \phi_D(n)} Q_{n}^{DR}+ \sum_{n \in \phi_{1}(n)} Q_{n}^{pv1} + \sum_{n \in \phi_{3}(n)} Q_{n}^{pv3}  \nonumber \\
    &+ \sum_{n \in \phi_{F}(n)} Q_{n}^{CC} = D_{n}^Q + \sum_{j \in \phi_n(n)} Q_{n,i}^{l}
\end{flalign}

The state of charge of each energy storage unit is defined using (\ref{SE_Charging2}). The minimum and maximum power limits of each battery are expressed in (\ref{SE_ramp}). The state of charge of each battery is constrained within a specified range using (\ref{SE_Capacity1}).
\begin{flalign}
   \label{SE_Charging2} & SE_{n} + P_{n}^S = PSE_{n} && \forall n \in \phi_{S} \\
    \label{SE_ramp}& R_{n}^L \le P_{n}^S \le R_{n}^H && \forall n \in \phi_{S} \\
    \label{SE_Capacity1}& C_{n}^{SL} \le  SE_{n} \le C_{n}^{SH}  && \forall n \in \phi_{S} 
\end{flalign}

Due to voltage fluctuations in the distribution system, controllable capacitor devices are considered in this paper. The controllable capacitor devices are modeled using (\ref{SVC_limit}).
\begin{flalign}
    \label{SVC_limit}& 0 \le Q_{n}^{CC} \le F_{n}^H && \forall n \in \phi_{F}
\end{flalign}

Demand response participants are incentivized based on their curtailed load. The amount of load curtailment which depends on customers’ behavior is restricted by maximum available DR capacity $DR_{m}D_{n,t}^P$, which is shown in (\ref{DR_P}).
\begin{flalign}
    \label{DR_P}& 0 \le P_{n}^{DR} \le DR_{m} D_{n}^P && \forall n\in \phi_D
\end{flalign}
where $DR_{m}$ is the maximum percentage of the load reduction that each customer can provide.

Three types of PV units are considered in this paper. For PV type 1, the reactive power output of each PV unit is controlled by an inverter and its active power is equal to the maximum available active power at each time, which are expressed using (\ref{PV1_Q}) and (\ref{PV1_P}), respectively.
\begin{flalign}
    \label{PV1_Q}& (Q_{n}^{pv1})^2 + (P_{n}^{pv1})^2 \le (C_{n}^{S1})^2  && \forall n\in \phi_{1} \\
    \label{PV1_P}& P_{n}^{pv1} = C_{n}^{P1} && \forall n\in \phi_{1}
\end{flalign}

The active power limit of each PV type 2 is presented in (\ref{PV2}). PV type 2 is not able to provide reactive power.
\begin{flalign}
    \label{PV2} 0\le &P_{n}^{pv2} \le C_{n}^{P2} && \forall n\in \phi_{2}
\end{flalign}

For PVs type 3, both active and reactive power can be controlled by their smart inverters. The active and reactive power constraints of each PV type 3 are defined as follows:
\begin{flalign}
    \label{PV3_P}& 0\le P_{n}^{pv3} \le C_{n}^{P3} && \forall n\in \phi_{3} \\
    \label{PV3_Q}& (P_{n}^{pv3})^2 + (Q_{n}^{pv3})^2 \le (C_{n}^{S3})^2
    && \forall n\in \phi_{3}
\end{flalign}

In order to keep voltage magnitude within a reasonable range, voltage limit constraints are modeled using (\ref{Voltage1}) (\ref{Voltage2}). It is assumed that the voltage magnitude of bus 1 which is connected to the substation is equal to 1 pu in (\ref{Voltage2}).
\begin{flalign}
    \label{Voltage1}& \frac {(V_{n}^{min})^2}{\sqrt{2}} \le U_{n} \le \frac{(V_{n}^{max})^2}{\sqrt{2}} && n \in \phi_{b}, n\neq 1 \\
    \label{Voltage2}& U_{n} = \frac{1}{\sqrt{2}} && n=1
\end{flalign}
where $V_{n}^{min}$ and $V_{n}^{max}$ are the minimum and maximum voltage limits.

\section{Proposed Two-stage Chance-Constrained Algorithm}

\begin{figure}
\centerline{\includegraphics[width=3.5in]{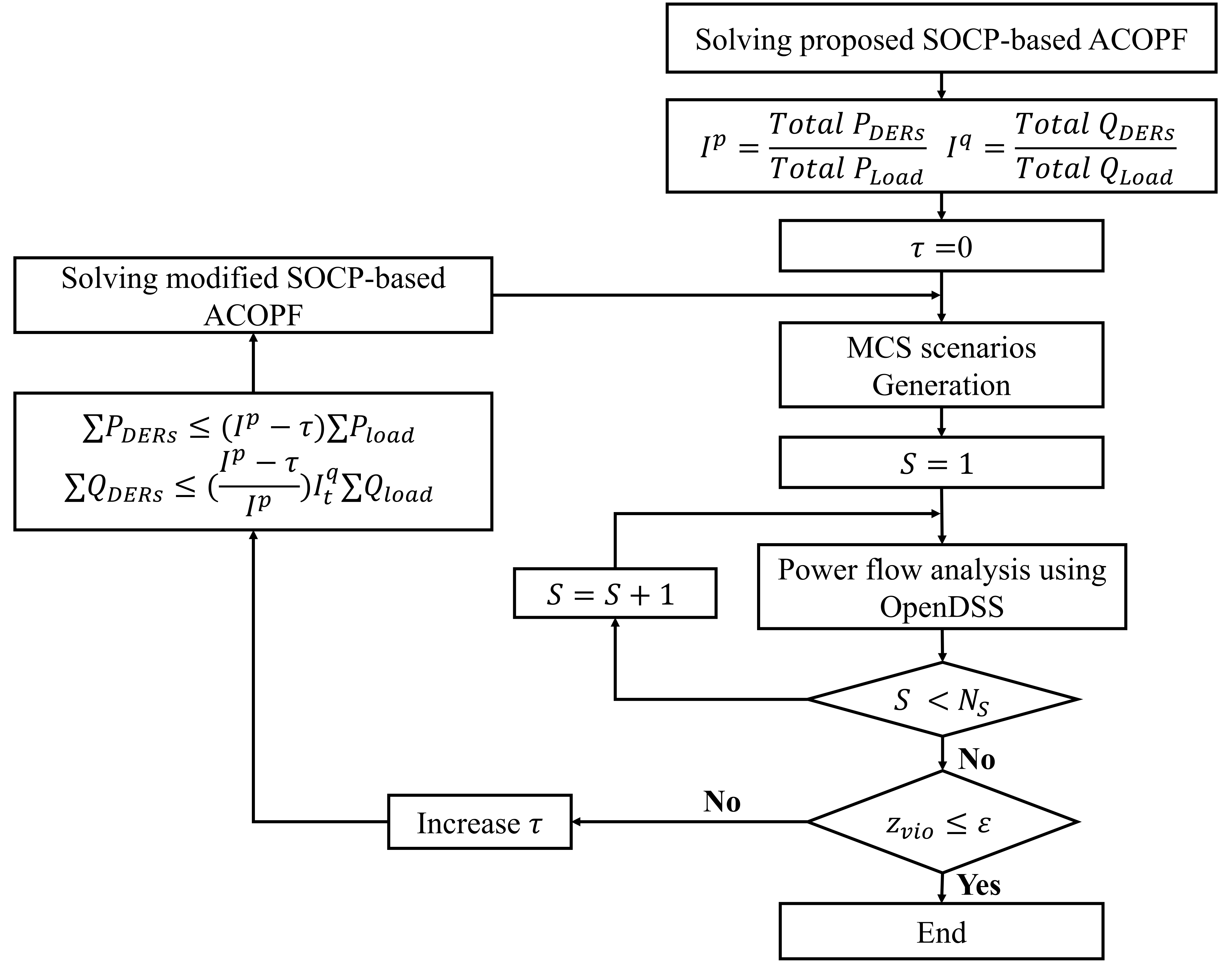}}
\caption{Two-stage Chance-Constrained SOCP-based ACOPF Model.}
\label{Figure50}
\end{figure}

Demand response affected by customers' behavior and PV units affected by weather conditions introduce uncertainties in distribution system operation. Modeling such uncertainties with increasing penetration of DERs may increase the operation cost and jeopardize the reliability of the distribution system. In this regard, in order to improve the economic efficiency and reliability of the system, a two-stage chance-constrained convex ACOPF model is proposed in this paper. The flowchart of the proposed method is shown in Fig. \ref{Figure50}.

In this paper, two indices are introduced to obtain the active and reactive participation ratio of DERs i.e., $I^p$ and $I^q$, which are the proportion of active and reactive power of total dispatched DERs to total active and reactive load of the system, respectively. In the first stage, the proposed SOCP-based ACOPF problem is solved with forecasted values of DERs, and active and reactive participation ratio of DERs are calculated. In the second stage, the Monte Carlo simulation is utilized to model uncertainties of DR resources and PV units. Also, a probability violation index is introduced in (\ref{zvio}), which represents the percentage of the scenario violating the allowed compensated power. The chance constraint is defined on the compensated power from the bulk grid. The compensated power is equal to the mismatch between the obtained injected power from the grid in modified SOCP-based ACOPF problem and injected power in each scenario of MCS. In other words, such power is the required reserve for secure and reliable real-time operation of the distribution grid due to the uncertainty of DERs. In this stage, power flow analysis is conducted for each scenario to determine the probability violation index of the system.
\begin{align}
    \label{zvio}z_{vio} = \frac{No. \ of \ Scenarios \ Violating \ the \ Threshold}{Total \ No. \ of \ Scenarios}
\end{align}
A probability violation criterion is set to make a trade-off between scheduling more DERs and imposing a higher risk to the system. If $z_{vio}$ is greater than the probability violation $\varepsilon$, the algorithm will reduce the participation ratios of DERs using $\tau$. The coefficient $\tau$ is initialized to be 0. In this paper, the reactive participation ratio of DERs is reduced following the reduction of the active participation ratio using $\frac{I^p-\tau}{I^p}$. Then, the proposed modified SOCP-based ACOPF is solved with less proportion of DERs until the probability violation criterion is satisfied. 

The proposed modified SOCP-based ACOPF model includes objective function given in (\ref{Obj}) and constraints (\ref{line_P})-(\ref{PV1_Q}), (\ref{PV2})-(\ref{Voltage2}), and (\ref{Adjust_PV1})-(\ref{Adjust_Q}). Since the active power output of PV type 1 is equal to the maximum available active power at each time, the constraint (\ref{PV1_P}) should be modified as (\ref{Adjust_PV1}).
\begin{flalign}
    \label{Adjust_PV1}& P_{n}^{pv1} = \frac{I^p-\tau}{I^p} C_{n}^{P1} && \forall n\in \phi_{1}
\end{flalign}

The two constraints (\ref{Adjust_P}) and (\ref{Adjust_Q}) are added to limit the total scheduling active and reactive power of DERs following the decrease of participation ratios.

\begin{flalign}
    \label{Adjust_P}&\sum_{n \in \phi_D} P_{n}^{DR}+ \sum_{n \in \phi_{1}} P_{n}^{pv1} + \sum_{n \in \phi_{2}} P_{n}^{pv2} + \sum_{n \in \phi_{3}} P_{n}^{pv3} \nonumber \\
    & \qquad \qquad \qquad \qquad \qquad \qquad \le (I^p-\tau) D_{n}^P  &&
\end{flalign}
\begin{flalign}
    \label{Adjust_Q}&\sum_{n \in \phi_D} Q_{n}^{DR}+ \sum_{n \in \phi_{1}} Q_{n}^{pv1} + \sum_{n \in \phi_{3}} Q_{n}^{pv3} \le (\frac{I^p-\tau}{I^p}) I^q D_{n}^Q  &&
\end{flalign}

\section{Case Study}

The performance of the proposed chance-constrained SOCP-based OPF model is evaluated by testing it on a modified IEEE 33-bus distribution system with three controllable capacitor devices, four energy storage units, six demand response resources, and five PV units for each type \cite{Reference18}. For the studied period, the total active and reactive power demand are 3529 kW and 2185 kVAR, respectively. Simulations are carried out in Python interfaced with OpenDSS, which is an open software for simulating the distribution grid \cite{ReferenceOpenDSS}. The SOCP-based ACOPF and modified SOCP-based ACOPF are programmed in Python and solved by Cplex while power flow analysis is conducted by OpenDSS. The wholesale electricity price information and DERs' price are obtained from ISO New England and Salt River Project, respectively \cite{Reference19}-\cite{Reference20}. To analyze the reliability of the distribution grid under the uncertainty of DERs, 1000 Monte Carlo simulations are carried out. Since residential customers' behavior is different at each bus, the active power of demand response at each bus is sampled from a distinct Gaussian distribution. Since PV generation is reliant on weather conditions, two different Gaussian distributions are utilized to simulate the power generation of PV types 1, 2 and 3 under sunny and cloudy weather. Since the inverters of PV type 1 units can control the reactive power and those of PV type 3 units can control both active and reactive power, it is not precise to model PV generation forecast error for these units based on active nor reactive power. In this regard, apparent power generation is sampled for the PV types 1 and 3, and active power generation is sampled for the PV types 2. 

\begin{figure}
\centerline{\includegraphics[width=3.5in]{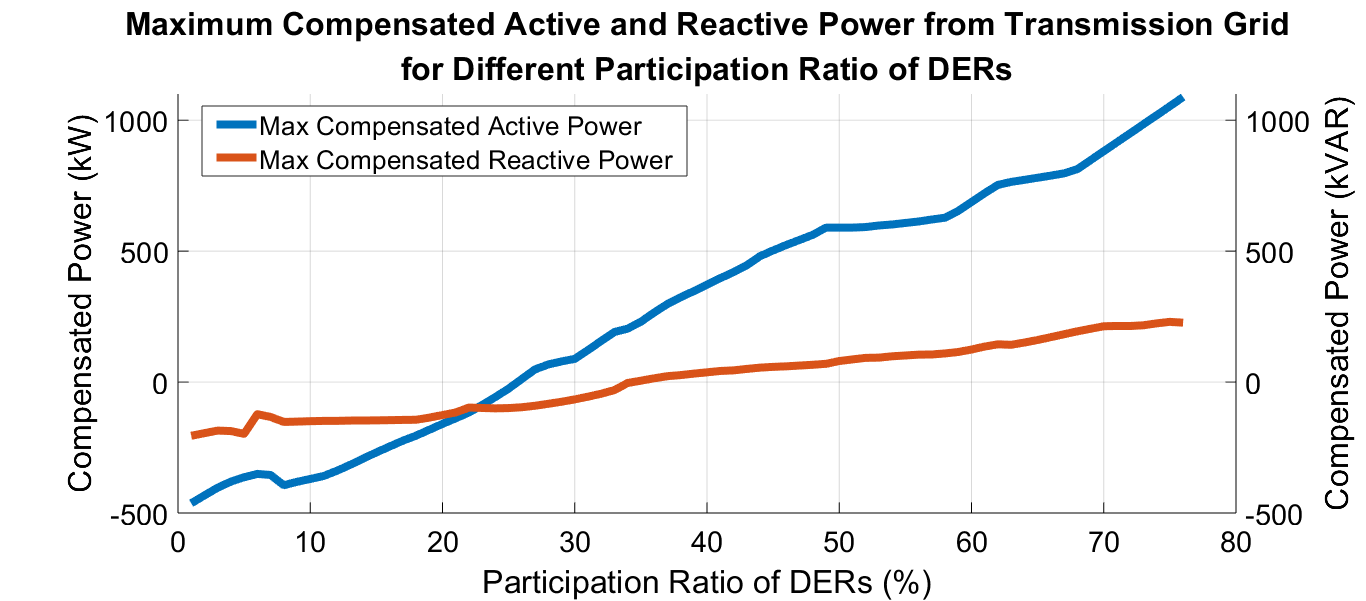}}
\caption{Compensated Injected Active and Reactive Power from Grid for Different Participation Ratio of DERs}
\label{Figure54}
\end{figure}

The increasing proportion of DERs is corresponding to the increasing uncertainty in the distribution grid. The relationship between maximum compensated active and reactive power among all MCS scenarios and the participation ratio of dispatched DERs is shown in Fig. \ref{Figure54}. This figure shows that due to the uncertain nature of DERs, the maximum compensated active and reactive power are increasing, as the proportion of DERs is increased. The maximum compensated power can also be considered as the reserves from the bulk grid. The higher the uncertainty in the distribution grid, the more reserves from the bulk grid the substation will tend to require in real-time operation.

\begin{figure}
\centerline{\includegraphics[width=3.5in]{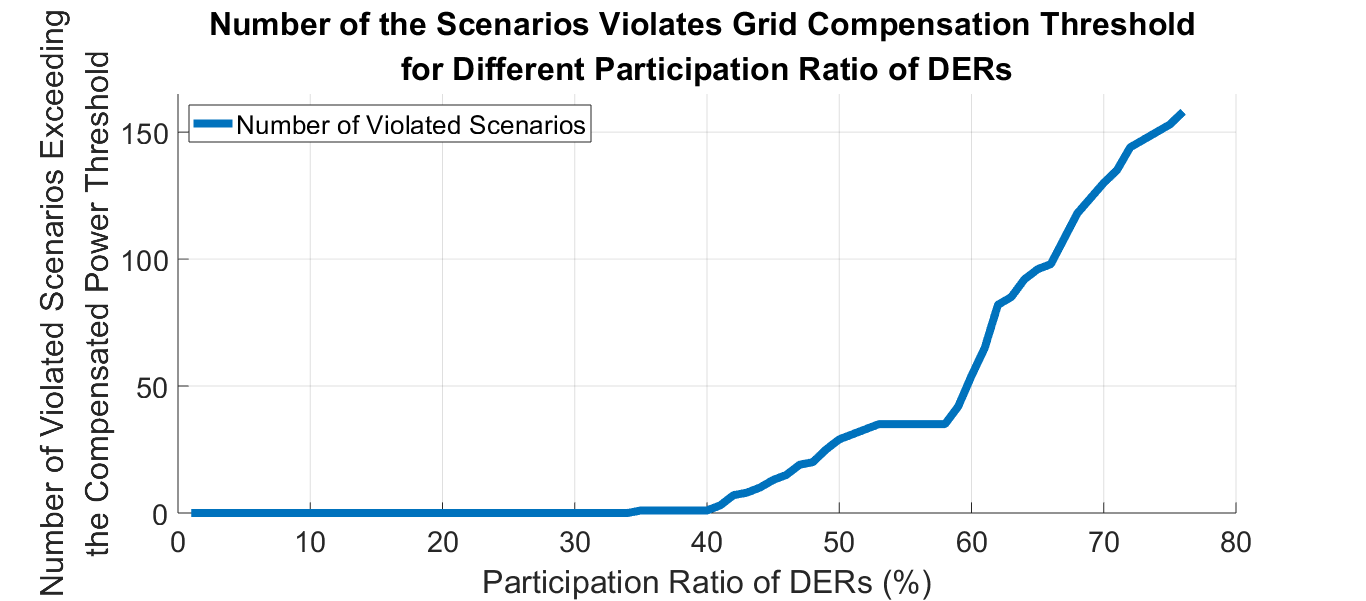}}
\caption{Number of Violated Scenarios with the 200 kW Compensated Power Threshold}
\label{Figure56}
\end{figure}

Since the reserves from the transmission grid may be limited, compensating more power from the transmission grid may bring troubles to the real-time operation of the system and undermine the reliability of the system. Moreover, since the real-time electricity price is much higher than the day ahead wholesale price, purchasing more power from the real-time market can affect customers and the utility's benefit. A threshold for compensated power is set to be 200 kW for active power. The number of scenarios is recorded for different participation ratios of DERs if the scenario violates the threshold and the results are exhibited in Fig. \ref{Figure56}. The results show that the number of scenarios that violate the compensated power threshold is increasing, as the participation ratios of DERs are increased. It implies that scheduling more DERs can result in requiring more compensated power from the bulk grid.

Different compensated power thresholds are modeled to evaluate their impacts on the dispatch of DERs, which are presented in Table \ref{Table1}. The probability violation $\varepsilon$ is fixed to be $5\%$. The compensated power threshold is changing from 100 kW to 600 kW. The Table \ref{Table1} shows the larger the compensated power threshold is, the more active and reactive power of DERs are dispatched. The results confirm that if there are enough reverses provided from the transmission grid for a determined system risk level, the more DERs will be scheduled. Since the DERs' price is cheaper than the wholesale price in the studied period, the cost of the system is decreased, if the participation ratios of DERs is increased. When the system is operated via a more conservative strategy i.e., less compensated power threshold, the proportion of dispatched DERs are reduced, and the cost of the system is increased.

\begin{table}[htbp]
  \centering
  \caption{Dispatched DERs for different compensated power threshold with 5$\%$ probability violation}
    \begin{tabular}{cccc}
    \toprule
    \multicolumn{1}{p{6.375em}}{\centering Compensated Power Threshold (kW)} & \multicolumn{1}{p{6.375em}}{\centering Participation Ratio of Active Power of DERs (\%)} & \multicolumn{1}{p{6.375em}}{\centering Participation Ratio of Reactive Power of DERs (\%)} & \multicolumn{1}{p{6.375em}}{\centering Operating Cost (\$)} \\
    \midrule
    \midrule
    100   & 50\%  & 20\%  & 136 \\
    \midrule
    200   & 59\%  & 24\%  & 134 \\
    \midrule
    300   & 63\%  & 26\%  & 133 \\
    \midrule
    400   & 68\%  & 28\%  & 132 \\
    \midrule
    500   & 70\%  & 29\%  & 132 \\
    \midrule
    600   & 73\%  & 30\%  & 132 \\
    \bottomrule
    \end{tabular}%
  \label{Table1}%
\end{table}%

The impact of probability violation is shown in Table \ref{Table2} with the fixed compensated power threshold which is equal to 200 kW for active power. Result shows that the proportion of dispatched active and reactive power of DERs is increasing as the probability violation increases. This implies that the more DERs tend to be scheduled if the operator can tolerate more violations. The small probability violation $\varepsilon$ value refers to a conservative operation strategy, while the large probability violation value implies an optimistic operation strategy. The more conservative the operator is, the less DERs power will be dispatched. Therefore, a conservative policy can lead to a higher system operating cost with less uncertainty in the distribution grid.

\begin{table}[htbp]
  \centering
  \caption{Dispatched DERs for Different probability violation with 200 kW Compensated Power Threshold}
    \begin{tabular}{cccc}
    \toprule
    \multicolumn{1}{p{6.375em}}{\centering Probability Violation} & \multicolumn{1}{p{6.375em}}{\centering Participation Ratio of Active Power of DERs (\%)} & \multicolumn{1}{p{6.375em}}{\centering Participation Ratio of Reactive Power of DERs (\%)} & \multicolumn{1}{p{6.375em}}{\centering Operating Cost (\$)} \\
    \midrule
    \midrule
    0.01  & 43\%  & 18\%  & 137 \\
    \midrule
    0.03  & 46\%  & 19\%  & 136 \\
    \midrule
    0.05  & 59\%  & 24\%  & 134 \\
    \midrule
    0.07  & 61\%  & 25\%  & 133 \\
    \midrule
    0.09  & 63\%  & 26\%  & 133 \\
    \midrule
    0.11  & 67\%  & 27\%  & 132 \\
    \bottomrule
    \end{tabular}%
  \label{Table2}%
\end{table}%

\section{Conclusion}

This paper deals with the uncertainty challenges of DERs in the balanced distribution system by proposing a two-stage chance-constrained convex ACOPF algorithm. In the first stage, a convex SOCP-based ACOPF model is proposed to obtain the optimal dispatch and participation ratio of DERs in the system. In the second stage, scenarios generated by Monte Carlo Simulation are utilized to determine the probability violation index, and the modified SOCP-based ACOPF model is applied to satisfy the probability violation criterion. The proposed chance-constrained method can handle the uncertainty of DERs, and guarantee the risk of the system is within the determined value. If distribution system operator determines a specific probability violation index, increasing participation ratios of DERs in the distribution system will lead to more compensated injected power from the bulk grid and less operating cost. However, the compensated power from the bulk grid may be limited in real-time operation, thus requiring more power from the grid can jeopardize the reliability of the distribution system. Also, with the increasing probability violation criterion, the proportion of dispatched active and reactive power of DERs is increased, and as a result, the economic efficiency of the distribution system will be enhanced while uncertainty in the distribution grid is increased. Therefore, the operator should make a trade-off between scheduling more DERs and imposing a higher risk and uncertainty to the distribution system. The proposed method can be extended to the three phases unbalanced network since ACOPF in the unbalanced network can also be convexified and formulated by utilizing the SOCP-based technique. The further study of the unbalanced system will be conducted in future work.

\end{document}